\documentclass{jltp}
\usepackage{graphicx}
\usepackage{psfig,graphics}
\usepackage{thumbpdf}
\usepackage{color}
\usepackage{amssymb}
\usepackage{bookmath}
\title{Domain Walls in Superfluid \Heb}
\author{A. Vorontsov and J. A. Sauls}
\address{Department of Physics and Astronomy, \\
         Northwestern University, Evanston, IL 60208, USA
    \\ {\usefont{OT1}{cmr}{m}{n} E-mail: anton@snowmass.phys.northwestern.edu}
    }
\runninghead{A. Vorontsov and J. A. Sauls}{Domain Walls in \Heb}
\begin{document}
\maketitle
\vspace{0.5cm}
\begin{abstract}
We consider domain walls between regions of superfluid \Heb\ in which
one component of the order parameter has the opposite sign in the two
regions far from one another. We report calculations of the order
parameter profile and the free energy for two types of domain wall, and
discuss how these structures are relevant to superfluid \He\ confined
between two surfaces.

\end{abstract}

The order parameter of superfluid \He\ is suppressed near an impurity or in
the vicinity of walls and interfaces.\cite{amb75,zha87,thu92,nag98} This
happens because the order parameter breaks the rotational
symmetry of the normal phase, and because particle-hole coherence is
destroyed when quasiparticles, which move in the presence of the order
parameter, are scattered from one point on the Fermi surface to another.
Here we consider a related problem in which the order parameter varies
strongly in space, so that quasiparticles move in this spatially varying
order parameter even when they propagate along straight trajectories.
Consider an order parameter that varies along some axis and has one bulk
solution on the far left, and another possible bulk solution on far right.
The region separating these two degenerate bulk solutions is the domain
wall. Since the order parameter is different on the two sides
the superfluid will be strongly deformed in the vicinity of the domain
wall. Although domain walls are not energetically favorable in bulk \He, we
will argue that inhomogeneous structures formed from domain
walls may be the favorable solutions for \He\ in confined geometries.

To determine the order parameter and to calculate the corresponding
free energy of a domain wall we use the quasiclassical theory.\cite{ser83}
We solve the Eilenberger transport equation for the quasiclassical
propagator, $\whg(\hat{\vp},\vR;\varepsilon_m)$, in the Matsubara formalism,
\be
[ i\varepsilon_m \widehat{\tau}_3 - \whDelta , \whg] + i\vv_f\grad \whg = 0
\label{eq:eil}
\,,
\ee
where $\varepsilon_m=(2m+1)\pi T$, $\whDelta$ is the $4\times 4$ matrix order
parameter and $\vv_f=v_f\hat{\vp}$ is the Fermi velocity corresponding to the
point $\vp_f$ on the Fermi surface. In addition, the physical solutions of the
quasiclassical propagator satisfy Eilenberger's normalization condition, $\whg^2
= -\pi^2 \widehat{1}$. It is convenient to express the propagator and order
parameter in particle-hole and spin space as,
\be
\whg = \left( \begin{array}{cc} g & \vf \cdot(i\vsigma \sigma_y) \\
(i\sigma_y \vsigma)\cdot\vf' & -g\end{array} \right) \,, \qquad
\whDelta = \left( \begin{array}{cc} 0 & (i\vsigma \sigma_y)\cdot\vDelta \\
(i\sigma_y \vsigma)\cdot\vDelta^*  & 0 \end{array} \right)  \,.
\ee
The vector in spin space, $\vDelta(\hat{\vp},\vR)$, depends on the spatial
position, $\vR$, and the direction of the momentum, $\hat{\vp}$. For pure
p-wave pairing the order parameter is traditionally written in terms of the
spin (index $\alpha$) and orbital (index $i$) matrix, $A_{\alpha i}(\vR)$,
\be
\Delta_\alpha(\hat{\vp},\vR) = \sum_{i=1}^{3} \, A_{\alpha i}(\vR) \, \hat{p}_i \,.
\ee
The gap equations for the order parameter components,
\bea
\lefteqn{
A_{\alpha i}(\vR) \ln {T\over T_c} = 3
\, T\sum_{\varepsilon_m>0}^{\omega_c} \,
\int \dangle{p} \, \hat{p}_i \, \times
}&&
\nonumber \\
&&\times \, \left( \vf(\hat{\vp}, \vR; \varepsilon_m) +\vf'(\hat{\vp}, \vR; \varepsilon_m)^* -
2 \pi {A_{\alpha i}(\vR) \hat{p}_i \over \varepsilon_m} \right) \,,
\eea
must be solved self-consistently with the transport equation.

Let us choose the $x$-axis to be the axis along which one of the
components of the order parameter is changing. The
domain wall is in the $yz$-plane centered at $x=0$. We assume the
order parameter to be translationally invariant in the $yz$-plane.
Reflection symmetries under $y\to\, -y$ and $z\to\, -z$
allow us to write the order parameter in a simple diagonal form,
\be
A_{\alpha i} =
\left(
\begin{array}{ccc}
A_{xx} & 0 & 0 \\
0 & A_{yy}& 0 \\
0 & 0 & A_{zz}
\end{array}
\right)
\equiv
\left(
\begin{array}{ccc}
\Delta_\perp & 0 & 0 \\
0 & \Delta_{1\, \parallel} & 0 \\
0 & 0 & \Delta_{2\, \parallel}
\end{array}
\right)
\,,
\ee
or equivalently in vector form,
\be
\vDelta(\vR,\hat{\vp}) = (\Delta_\perp(\vR) \hat{p}_x,
\Delta_{1\, \parallel}(\vR) \hat{p}_y, \Delta_{2\, \parallel}(\vR) \hat{p}_z)
\,,
\ee
where the symbol $\parallel$ ($\perp$) refers to momentum directions
parallel (perpendicular) to the domain wall.

We consider two basic configurations of domain wall. The bulk phase of
superfluid \He\ at zero pressure is the B-phase with the order parameter
$\vDelta_0(\hat{\vp}) = (\Delta_0 \hat{p}_x, \Delta_0 \hat{p}_y, \Delta_0
\hat{p}_z)$. Other possible bulk solutions differ from this one by a sign
change of one of the components.
We consider the first domain wall configuration to be one with an
order parameter
$\vDelta(\hat{\vp})_L=(-\Delta_0 \hat{p}_x, \Delta_0 \hat{p}_y, \Delta_0 \hat{p}_z)$
on the far left, and
$\vDelta(\hat{\vp})_R=(+\Delta_0 \hat{p}_x, \Delta_0 \hat{p}_y, \Delta_0 \hat{p}_z)$
on the far right. We refer to this configuration as the
``perpendicular'' domain wall because it is $\Delta_\perp$ that changes sign
across the domain wall. This configuration is equivalent to the problem of
determining the order
parameter structure near a specularly reflecting surface positioned at $x=0$.
In the case of a specular surface a quasiparticle with momentum
$(p_x, p_y, p_z)$ is scattered to a state with momentum $(-p_x, p_y, p_z)$.
As a result these specularly reflected quasiparticles move in the same order
parameter field as quasiparticles moving along straight trajectories through
a perpendicular domain wall.
The second domain wall configuration we consider has
$\vDelta(\hat{\vp})_L = (\Delta_0 \hat{p}_x, \Delta_0 \hat{p}_y, -\Delta_0 \hat{p}_z)$
on the left and
$\vDelta(\hat{\vp})_R = (\Delta_0 \hat{p}_x, \Delta_0 \hat{p}_y, +\Delta_0 \hat{p}_z)$
on the right side, and we refer to this structure as a
``parallel'' domain wall. This structure is not related
any order parameter structure defined by a reflecting surface.

\begin{figure}
\centerline{
\includegraphics[height=6cm]{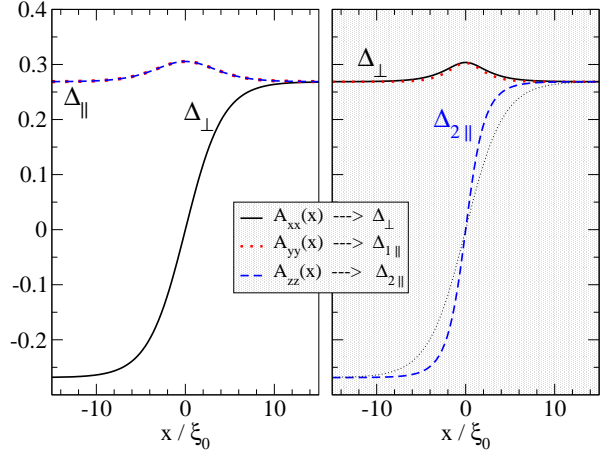}
}
\caption{The order parameter profile across the domain wall
($\Delta$ is in units of $2\pi k_B T_c$, $\xi_0 = \hbar v_f/2\pi k_B T_c$).
Left panel: change in $\Delta_\perp$ across the wall,
$A_{xx}(-x)=-A_{xx}(x)$.
In this case $\Delta_{1\,\parallel}=\Delta_{2\,\parallel} = \Delta_{\parallel}$.
Right panel: $\Delta_{2\,\parallel}$ changes sign, $A_{zz}(-x)=-A_{zz}(x)$, and
$\Delta_{1\,\parallel}$ is slightly different from $\Delta_\perp$.
For comparison the thin dotted line is $\Delta_\perp$ from the left panel. The
parallel domain wall is thinner than the ``perpendicular'' domain wall
by a factor of about $1.8$.
}
\label{fig:op}
\end{figure}

The self-consistent solutions for the two order parameter configurations are
shown in Fig. \ref{fig:op}. We see when $\Delta_\parallel$ changes sign (right
panel) the domain wall is approximately twice as narrow as than the domain wall
in which $\Delta_\perp$ changes sign. This means that the deformation of the
superfluid order parameter is weaker for the case of the ``parallel'' domain
wall. The difference in the order parameter structures between these two domain
walls is qualitatively explained by difference in the number of trajectories
that are pairbreaking. In the ``perpendicular'' configuration the pairbreaking
trajectories are predominantly perpendicular to the domain wall. In the case of
the ``parallel'' domain wall trajectories that are nearly parallel to domain
wall are most pairbreaking. However, these trajectories only weakly connect the
two sides of the domain wall. As a result the profile of the order parameter is
narrower than that of the perpendicular domain wall.

We explore this feature further by computing the free energy for each given
order parameter profile. The free energy is calculated by introducing the
auxiliary propagator $\whg_\lambda$ that satisfies the transport equation,

\be\label{eq:a2-auxiliary_transport}
\left[i\varepsilon_m\widehat{\tau}_3-\lambda\,\whDelta,\whg_\lambda\right]
+i\vv_f\cdot\grad\whg_\lambda = 0
\,,
\ee
with $0\le\lambda\le 1$. The free energy in the weak coupling limit
can be expressed as an integral over the
variable coupling constant, $\lambda$,\cite{vor03c}
\be
\Del\Omega =
\onehalf\int_0^1\,d\lambda\,
\mbox{Sp}'\left\{\whDelta\,\left(\whg_\lambda-\onehalf\whg\right)\right\}
\equiv \int d^3 R \; \Del F(\vR)
\,.
\label{eq:a2-FE}
\ee
We define the free energy density, $\Del F(\vR)$, from the definition of the
trace, $\mbox{Sp}'$, which stands for summation and integration over all indices and
variables, e.g.
\be
\mbox{Sp}'\left\{\whDelta\,\whg\right\}\equiv
\int d^3 R \left\{ T\sum_{\varepsilon_m}
N_f \, \int\frac{d\Omega_{\hat\vp}}{4\pi}
\mbox{Tr}_4
\big(\whDelta(\hat\vp,\vR)\whg(\hat\vp,\vR;\varepsilon_m)\big) \right\} \,.
\nonumber
\ee

\begin{figure}
\centerline{\includegraphics[height=6cm]{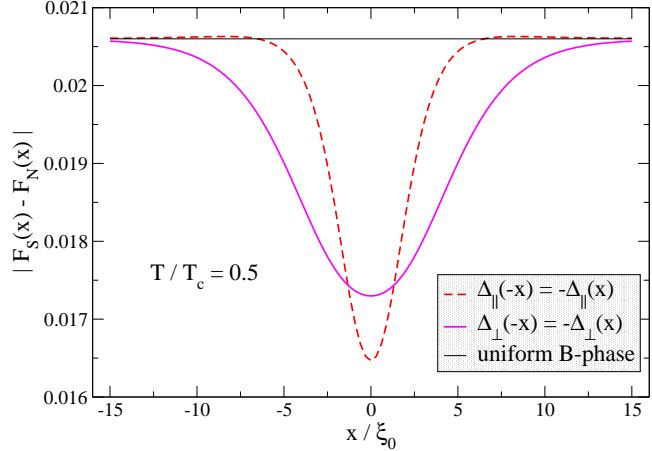}}
\caption{Reduction of the condensation energy near a domain wall
[in units $N_f(2\pi k_B T_c)^2$].
The magnitude of the free energy density for the ``parallel'' configuration is suppressed
more around $x=0$ due to the larger gradient energy terms, but the integrated
suppression is less than that for the ``perpendicular'' configuration.}
\label{fig:fe}
\end{figure}

The free energy densities for the two domain walls are shown in Fig.
\ref{fig:fe}. The change in $\Delta_{2\,\parallel}$ takes place over a
shorter distance than that of $\Delta_\perp$. Thus, the former
configuration has larger gradient energy. Nevertheless, this is
compensated by the spatial extent of ``perpendicular'' domain wall with
the result that the net energy cost of the ``parallel'' domain wall is
less than that of the "perpendicular'' domain wall. This result has
implications for the structure of the order parameter and the stability
of new phases of superfluid \He\ in confined geometry.

\begin{figure}
\centerline{\includegraphics[height=3cm]{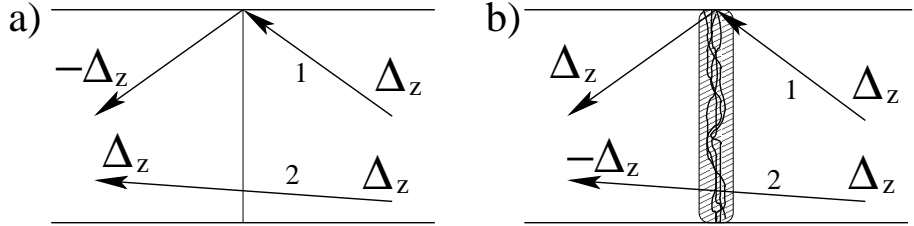}}
\caption{Pairbreaking trajectories with and without a domain wall in the \Heb\ film.
a) no domain wall, b) $A_{zz}(x)$ changes sign from left to right.}
\label{fig:film}
\end{figure}

We demonstrate this point by the following argument. Fig. \ref{fig:film} shows a
slab of \Heb\ and two types of quasiparticle trajectories. Trajectory $1$
undergoes a reflection from a boundary, while trajectory $2$ is a straight
trajectory in the region of interest. In case (a) shown in Fig. \ref{fig:film} the
order parameter component $A_{zz}$ has the same sign to the left and right of the
vertical line. Thus, quasiparticles moving on trajectory $1$ propagate in a
changing order parameter field, $\Delta_z \to-\Delta_z$, which is equivalent to
the ``perpendicular'' domain wall. However, quasiparticles moving along trajectory
$2$ do not experience a change in the sign of the order parameter.

Case (b) shown in Fig. \ref{fig:film} is quite different. In this case we
assume a domain wall in which $A_{zz}$ changes sign on the two sides. Now
quasiparticles moving on trajectory $1$ do not experience a sign change
because the change in the momentum, $\hat{p}_z \to \, -\hat{p}_z$, is
compensated by the sign change in $A_{zz}$. Thus, there is no contribution
to the order parameter deformation from this trajectory. However, this
reduction in surface pairbreaking occurs at the cost of pairbreaking
introduced trajectories of the second type. Quasiparticles moving on
trajectory $2$ propagate through an order parameter field with a sign
change. But since these type of trajectories lead to a domain wall of the
``parallel'' type, the domain wall structure may be favorable to the
uniform solution in the thin film shown in case (a).

This observation about the possible stability of domain wall structures depends
on the thickness of the slab or film since for very thick films it is clearly
unfavorable to have domain walls in the plane of the film. However, for thinner
films or slabs inhomogeneous structures related to parallel domains are possible.
A full quasiclassical calculation of the structure and free energy for thin films
of superfluid \He\ confirms this observation based on the domain wall
calculations. We have found that associated with the domain wall structures new
order parameter components develop, and that there are periodic structures formed
from domain walls along the plane of the film that are energetically favorable
compared to an order parameter that is translationally invariant in the plane of
the film. We find a range of film thickness, $10\xi_0 - 12 \xi_0$, for which
these inhomogeneous states are favored.
The details of this calculation will be published elsewhere.

We thank Tomas L\"ofwander and Erhai Zhao for useful comments on this work.


\end{document}